\documentclass[12pt]{iopart}

\usepackage{graphicx}
\usepackage{bm}
\newcommand {\bnabla} {\mbox{\boldmath$\nabla$}}

\newcommand {\bell} {\mbox{\boldmath$\ell$}}

\newcommand {\bcalE} {\mbox{\boldmath${\cal E}$}}   

\begin{document} 

\title[A comprehensive continuum theory of structured liquids]{A comprehensive continuum theory\\ of structured liquids}

\author{R. Blossey$^1$ and R. Podgornik$^{2,3,4}$} 

\address{$^1$ University of Lille, Unit\'e de Glycobiologie Structurale et Fonctionnelle (UGSF) CNRS UMR8576, 
59000, Lille, France\\ 
$^2$ School of Physical Sciences and Kavli Institute for Theoretical Sciences, University of Chinese Academy of Sciences, Beijing 100049, China\\
$^3$ CAS Key Laboratory of Soft Matter Physics, Institute of Physics, Chinese Academy of Sciences, Beijing 100190, China and 
Wenzhou Institute of the University of Chinese Academy of Sciences, Wenzhou, Zhejiang 325000, China\\
$^4$ Also at: Department of Physics, Faculty of Mathematics and Physics, University of Ljubljana, Ljubljana, Slovenia}
\ead{ralf.blossey@univ-lille.fr; podgornikrudolf@ucas.ac.cn}  

\begin{abstract}
We develop a comprehensive continuum model capable of treating both electrostatic and structural interactions in liquid dielectrics. 
Starting from a two-order parameter description in terms of charge density and polarization, we derive a field-theoretic model generalizing previous theories.
Our theory explicitly includes electrostatic and structural interactions in the bulk of the liquid and allows for polarization charges within a Drude model. 
In particular we provide a detailed description of the boundary conditions which include the charge regulation mechanism and surface polarization, which is
explained both in general terms and analyzed for an exemplary model case. Future applications of our theory to predict and validate experimental results are 
outlined.
\end{abstract}

\maketitle

\section{Introduction}\label{sec1}

The continuous development and improvement of modern experimental techniques is pushing the resolution limits 
in soft matter systems continuously further down towards atomistic regimes. As an example one can take the
advances made with the application of different varieties of atomic-force microscopy (AFM) in liquids. Both conventional methods 
and the recently developed 3D-AFM technique \cite{Fukuma2010,Kimura2010,Herruzo2013,Fukuma2018,vanLin2019,Klaassen2022} 
meanwhile allow access to unprecedented molecular and atomistic detail
of liquids, notably  aqueous ionic solutions, near hydrophilic and hydrophobic surfaces. 

These successes, in turn, challenge theory. The parallel advances made in molecular computation, in combination with machine learning approaches, seem to render purely theoretical approaches obsolete: theorists may want to lay down their original weapons and declare defeat. This view was recently expressed in the work of Mugele and collaborators. They notice that it is difficult or perhaps impossible to decouple the total monotonically decaying force into various well-defined separate contributions like DLVO and/or non-DLVO interactions, ``because one cannot develop a universal Poisson-Boltzmann theory accounting for all nonelectrostatic effects (ionic chemical nature, size, charge, polarizability, and solvation)'' \cite{Klaassen2022}. 

In this paper we take up this challenge by developing a comprehensive field-theoretic continuum model extending standard Poisson-Boltzmann-type theories. The starting point of our analysis is our recently developed field theory
of structured dielectrics \cite{Blossey2022}, in which the bulk theory was developed based on a two-order
parameter description in terms of charge density and polarization. In the development of this theory we were
guided by the Onsager-Dupuis theory of the dielectric properties of ice \cite{Onsager1960,Onsager1962,Gruen1983,Gruen1983-2}. For a recent perspective on how our work \cite{Blossey2022} can be placed in the context of earlier theories of polarization, see \cite{Blossey2022-2}. 

In order to bring this approach close to experiment, in this work we extend the basic theory in several
respects. First, for the bulk system, we go beyond the previous formulation by explicit allowance for 
polarizability of both the molecular solvent as well as the ionic solutes, which we treat in terms of a Drude model. More importantly, however, we develop
a detailed formulation of the boundary conditions, overall missing from previous endeavours, allowing for surface polarization as well as for
charge regulation mechanisms. This part is the main novel contribution 
of the present paper. It is discussed both in general terms as well as elucidated for exemplary cases.
In closing, we discuss experimental systems of interest to which our theory can be applied in the future. 

\section{A continuum model for a polarizable dipole-ion mixture}\label{sec2}

In our previous work we considered a structured liquid dielectric described by charge density and solvent
polarization in a harmonic approximation and formulated the corresponding field theory for a bulk system \cite{Blossey2022}.  
Here, we first revisit the formulation of this theory by including the polarizability of the solvent dipoles and solute ions, the non-electrostatic 
dipole-dipole interaction and the charge-dipole hydration coupling. In the subsequent step, we derive a general expression for the surface 
free energy and the associated boundary conditions, which completes the development of our formalism  and prepares our theory for applications
to a variety of experimental systems.

\subsection{
Bulk equations}

The physical components of the system we study are the solvent molecules and the solute electrolyte cations and anions. 
The polarizability of the constituents is implemented {\sl via} harmonically connected Drude charges; this is fully described in Appendix A. 
Our approach begins with the identification of the appropriate {\sl order parameters}; for a motivation, see, e.g., Ref. \cite{Podgornik2021}. 
For the simple electrolyte component we can define the {\sl cation} and {\sl  anion density fields }
\\
\begin{eqnarray}
&& \hat{\rho}_{(\pm)}({\bf x}) \equiv  \pm e \sum_{(\pm)} \delta\left( {\bf x} - {\bf x}_n \right) + q_I \sum_{(\pm)} {\bell}_i\cdot  \bnabla~\delta({\bf x} - {\bf x}_i),
\end{eqnarray}
\\
where $e$ is the elementary charge of the salt ions and $q_I$ (assumed to be the same for both types of ions) are the Drude charges, with a fluctuating separation $\bell$  exhibiting an average of $\left< \bell_i\right> = 0$ and a variance of $\left< \bell_i\cdot\bell_i \right> = \left< \ell_i^2 \right> =\sigma_I^2$. The solvent molecule dipolar charge is now given as 
\\
\begin{eqnarray}
\hat\rho_{(N)}({\bf x}) = 
 &=& q_D \sum_{(N)} {\bell_i}\cdot  \bnabla~\delta({\bf x} - {\bf x}_i).
\end{eqnarray}
\\
For the solvent molecules the fluctuating distance between the two Drude charges, $\bell_i$, has an average of $\left< \bell_i\right> = s$ and a variance of $\left< \bell_i^2 \right> =\sigma_D^2$. 
Throughout the text we will use the same symbol for the ionic and polarization Drude fluctuating separation; the context and properties being clearly different.

The {\sl total charge density field} subject to Coulomb interactions is then given by the sum of the ionic charges and the divergence of the total polarization vector
\\
\begin{eqnarray}
 \hat{\rho}({\bf x}) &=& 
\pm e \sum_{(\pm)} \delta\left( {\bf x} - {\bf x}_i \right) + \bnabla\cdot \hat{\bf P}({\bf x}).
\label{cole3}
\end{eqnarray}
\\
with the total polarization given by 
\\
\begin{eqnarray}
\hat{\bf P}({\bf x}) = q_D \sum_{(N)} {\ell}_i \delta({\bf x} - {\bf x}_i) + q_I \sum_{(\pm)} {\bell}_i\delta({\bf x} - {\bf x}_i),
\label{cole3a}
\end{eqnarray}
\\
where we reiterate that $\left< \bell_i\right> = s$ for the solvent ($N$) and $\left< \bell_i\right> = 0$ for ions ($\pm$). 
The total Coulomb interaction energy has the standard form
\\
\begin{eqnarray}
{{\cal H}_C}  &=& {\textstyle\frac12} \int\!\!\int_V d{\bf x}~d{\bf x}' ~ {\hat\rho}({\bf x})u({\bf x} -{\bf x}') {\hat\rho}({\bf x}') 
\label{hamilt1}
\end{eqnarray}
\\
where the Coulomb kernel corresponds to the dielectric constant given solely by its non-configurational part, corresponding to the high-frequency dielectric constant, $\varepsilon_{\infty} = \varepsilon(\omega \longrightarrow \infty)$, {\sl i.e.}
\\
\begin{equation}
u  = \frac{1}{4\pi \varepsilon_{\infty}\varepsilon_0~ \vert {\bf x} -{\bf x}' \vert}\,, \quad \mbox{hence}\quad u^{-1}  = - \varepsilon_{\infty}\varepsilon_0~ \bnabla^2 \delta({\bf x} -{\bf x}'), 
\label{Coulomb}
\end{equation}
\vspace*{0.5cm} 
where $\varepsilon_{\infty}$ accounts for all the relaxation mechanisms at higher relaxation frequencies. Clearly, while the Coulomb potential presents a non-local coupling, its inverse is a purely local operator.

The non-electrostatic short-range interactions, denoted by a tilde symbol, act between local solvent dipoles and can be written in a quadratic approximation as 
\\
\begin{eqnarray}
{{\cal H}_{SR}}  =  {\textstyle\frac12} \int\!\!\int_V d{\bf x}~d{\bf x}' ~ \hat{\cal \bf P}_{i}({\bf x})\tilde u_{ij}({\bf x} -{\bf x}') \hat{\cal \bf P}_{j}({\bf x}'),
\label{hamilt0}
\end{eqnarray}
\\
where $\tilde u_{ij}({\bf x} -{\bf x}')$ is a short-range, non-electrostatic interaction potential. A minimal model expression for this potential is the fourth-order non-local derivative expression
\\
\begin{eqnarray}
\tilde u_{ij}({\bf x} -{\bf x}') & = & u_P(0) \Big( \delta_{ij} \delta({\bf x} -{\bf x}') + \xi^2~\bnabla'_j\bnabla_i \delta({\bf x} -{\bf x}') \nonumber \\
& + & \zeta^4 \bnabla'_k\bnabla'_j\bnabla_k\bnabla_i \delta({\bf x} -{\bf x}') \Big)  +  \dots  
\label{shortrange1}
\end{eqnarray}
\\
for the dipolar interactions, containing the correlation length $\xi$ and the structural length $\zeta $ of the solvent, respectively. This implies the quadratic form of the structural interaction free energy
\\
\begin{equation}
\hspace*{-0.8cm} {{\cal H}_{SR}}  = {\textstyle\frac12} u_P(0) \int_V d{\bf x} \Big( \hat{\cal\bf P}^2({\bf x}) + \xi^2 \left(\bnabla\cdot\hat{\cal\bf P}({\bf x})\right)^2 + \zeta^4 \left(\bnabla (\bnabla\cdot\hat{\cal\bf P}({\bf x}))\right)^2\Big). 
\label{hamilt0a2}
\end{equation}
\vspace*{0.5cm} 
There is no universality in the non-electrostatic interactions and they are invariably linked with different models of the liquid structure. 
Higher-order terms either in the polarization vector or its derivatives are also possible, {\sl e.g.}, close to an ordering transition of water dipoles where the polarization energy 
could be written in a Landau form consistent with presumed microscopic symmetries \cite{Maggs2006,Monet2021}.

Next we consider the hydration shell of the ions that corresponds to the coupling between the ion density and $\hat{\rho}_{(\pm)}({\bf x})$ and $\hat{\cal \bf P}({\bf x})$. To the lowest order this coupling 
can be written as
\\
\begin{equation}
    {\cal H}_{HY}  =   \int\!\!\int_V d{\bf x}d{\bf x}' ~ \hat{\rho}({\bf x})\tilde u({\bf x} -{\bf x}')~ \bnabla\cdot{\hat{\cal \bf P}}({\bf x}'), 
      \label{equcoupling}
 \end{equation}
\vspace*{0.5cm} 
where the potential $\tilde u({\bf x} -{\bf x}')$ is again a short-range, non-electrostatic potential that can be modeled as 
\\
\begin{equation}
\tilde u({\bf x} -{\bf x}') = \alpha ~\delta({\bf x} -{\bf x}') +  \dots \, .
\label{shortrange2}
\end{equation} 
\vspace*{0.5cm} 
We assumed that the hydration polarization for anions and cations is - apart from the direction - identical for
both. The implied hydration free energy then takes the form
\\
\begin{equation}
    {\cal H}_{HY}  =   \alpha \int_V d{\bf x} ~ \hat{\rho}({\bf x})~ \bnabla \cdot{\hat{\cal \bf P}}({\bf x}). 
      \label{equcoupling2}
 \end{equation}
\vspace*{0.5cm}
The total interaction energy equals the sum of the above three contributions, {\sl i.e.}, 
\\
\begin{equation}
{\cal H} = {\cal H}_C + {\cal H}_{SR} + {\cal H}_{HY}.
\end{equation} 
\vspace*{0.5cm} 

In the next step we introduce the two collective variables, ${\rho}({\bf x})$ and ${\cal \bf P}({\bf x})$, which imply also two auxiliary fields, 
${\bcalE}(x)$, $\phi(x)$, that on the mean-field level turn out to be the non-electrostatic part of the polarization vector and the electrostatic potential scalar.
On the saddle-point (mean-field level) the free energy can be cast into the form of a functional of the 
collective order parameters and auxiliary fields given by
\\
\begin{eqnarray}
\hspace*{-2cm} \beta {\cal F}[{\cal \bf P}({\bf x}), {\rho}({\bf x}); \bcalE({\bf x}), \phi({\bf x})] \equiv 
{\textstyle\frac12} \int\!\!\int_V d{\bf x} d{\bf x}' ~ {\cal \bf P}_{i}({\bf x})\tilde u_{ij}({\bf x} -{\bf x}') {\cal \bf P}_{j}({\bf x}') + \nonumber \\
{\textstyle\frac12} \int\!\!\int_V d{\bf x} d{\bf x}' ~ {\rho}({\bf x})u({\bf x} -{\bf x}') {\rho}({\bf x}') + 
\int\!\!\int_V d{\bf x} d{\bf x}' ~ {\rho}({\bf x})\tilde u({\bf x} -{\bf x}') \bnabla \cdot{{\cal \bf P}}({\bf x}') \nonumber \\
- \int_V d{\bf x}~ {\cal \bf P}_{i}({\bf x}) \bcalE_{i}({\bf x}) - \int_V d{\bf x} ~\rho({\bf x})\phi({\bf x}){ -V[\bcalE({\bf x}), \phi({\bf x})]}, \nonumber \\
 \label{maineq1}
\end{eqnarray}
where $V[\bcalE({\bf x}), \phi({\bf x})]$ is the one-particle partition function of the system in an external field with the Hamiltonian $\tilde{\cal H}^*$ of the form
\\
\begin{eqnarray}
&& \beta \tilde{\cal H}^*[\bcalE({\bf x}), \phi({\bf x})] =  \int_V d{\bf x} ~\hat{\cal \bf P}({\bf x})\cdot \bcalE({\bf x}) + \int_V d{\bf x} ~\hat{\rho}({\bf x}) \phi({\bf x}) \nonumber\\
&&~~~~~~~~~~~ =  q_D \sum_{(N)}   {\ell}_i ~\! \big( \bcalE({\bf x}_i)-\bnabla \phi({\bf x}_i) \big) 
+ q_I \sum_{(\pm)}   {\ell}_i ~\! \big( \bcalE({\bf x}_i)-\bnabla \phi({\bf x}_i) \big) + \nonumber \\
&&~~~~~~~~~~~~~~~ + \sum_{(+)} ~e \phi({\bf x}_i) -  \sum_{(-)} ~e \phi({\bf x}_i), 
\end{eqnarray}
\\
with vector and scalar (external) auxiliary fields. The one particle partition function is then obtained by taking the trace over all the particle degrees of freedom, that is
\\
\begin{eqnarray}
e^{-V[\bcalE({\bf x}), \phi({\bf x})]} = \left< e^{-\beta \tilde{\cal H}^*}\right>
\end{eqnarray}
\\
where the symbolic average stands for the trace over the particle coordinates and internal degrees of freedom, 
{\sl i.e.}, the extension $\ell_i$ and the orientation ${\bf n}_i$ for the Drude model description. 
In addition, this average can be taken either for a grand canonical ensemble or for a lattice gas ensemble, resulting in two different forms of the one-particle partition function $V[\bcalE({\bf x}), \phi({\bf x})]$. The one-particle partition function can be obtained explicitly for a mixture of a polarizable solvent - electrolyte solute (see Appendix B for details of the calculation), by first defining 
\\

\[
\hspace*{-2cm} \upsilon\left(\bcalE({\bf x}), \phi({\bf x})\right) \equiv  e^{\mu_D} 
\frac{\sinh{
\left(\beta q_D s~ \vert\bnabla\phi({\bf x}) - \bcalE({\bf x})\vert
\right)}}
{q_D \beta s~\vert\bnabla\phi({\bf x})-  \bcalE({\bf x})\vert} 
~ e^{{\textstyle\frac12}{(\beta\sigma q_D)^2}\left(\bnabla\phi({\bf x}) 
-  \bcalE({\bf x})\right)^2} 
\]
\vspace*{0.5cm} 
\begin{equation} 
+  2 e^{\mu_\pm}  \cosh{\beta e \phi({\bf x}) e^{{\textstyle\frac12}{(\beta\sigma q_I)^2}\left(\bnabla\phi({\bf x}) -  \bcalE({\bf x})\right)^2}}. 
\label{bgwfxewa1}
\end{equation}
\vspace*{0.5cm} 
In the grand canonical ensemble we have
\\
\begin{eqnarray}
V[\bcalE({\bf x}), \phi({\bf x})] = \int d^3{\bf r} ~\upsilon\left(\bcalE({\bf x}), \phi({\bf x})\right), 
\end{eqnarray}
\\
while the lattice gas ensemble leads to
\\
\begin{eqnarray}
V[\bcalE({\bf x}), \phi({\bf x})] = \frac{1}{a^3} \int d^3{\bf r} ~\ln{\upsilon\left(\bcalE({\bf x}), \phi({\bf x})\right)}\, .
\end{eqnarray}
\\
We assume that for a univalent electrolyte the chemical potentials satisfy $\mu_\pm = \mu_- = \mu_+$, 
and $a^3$ is the volume of a site in the lattice gas partition function. Apart from the contribution 
of the polarizability on the Drude model level, the expression matches the one derived before \cite{Blossey2022}. 
We also note that in the purely electrostatic case ($\bcalE = 0$, see below) and with equal polarizability 
for the solvent and solute species, the polarizability terms decouple from the rest of the free energy.

The above expressions differ from the case of the fixed dipole-ion mixture analyzed by Abrashkin et al. \cite{Abrashkin2007,Gongadze2014}, 
as well as from the polarizable dipole-ion mixture analyzed by Buyukdagli et al. \cite{Buyukdagli2013}, or indeed from the polarizable ion case as 
formulated by Demery et al. \cite{Netz2001,Demery2012}, or the work of L\'evy et al. \cite{Levy2013}, where a single scalar order parameter 
has been introduced to characterize the solvent and solute density. In our formulation there are, indeed, two crucial  differences: the inclusion of the polarizability 
terms for the solvent molecules as well as the solute ions, and the dipolar orientational field, $\bnabla\phi({\bf x}) -  \bcalE({\bf x})$, which is given by the sum 
of the non-electrostatic component of polarization and the electrostatic field, {\sl i.e.},  accounting for electrostatic as well as non-electrostatic orientational ordering, 
as we show explicitly in what follows.

Inserting the non-local potentials from Eqs. (\ref{Coulomb}), (\ref{shortrange1}), (\ref{shortrange2}) into the free energy allows us first to write down the equivalent 
mean-field form of the free energy, Eq. (\ref{maineq1}), as 
\\
\begin{eqnarray}
\hspace*{-1.5cm}\beta {\cal F}_V[{\cal \bf P}({\bf x}); \bcalE({\bf x}), \phi({\bf x})] \equiv \\  
\hspace*{-1.5cm}{\textstyle\frac12} u_P(0) \int_V d{\bf x} \left({\cal\bf P}^2({\bf x}) + \xi^2 \left(\bnabla\cdot{\cal\bf P}({\bf x})\right)^2 
+ \zeta^4 \left(\bnabla (\bnabla\cdot{\cal\bf P}({\bf x}))\right)^2\right) \nonumber\\
\hspace*{-1.5cm} - {\textstyle\frac12}\varepsilon_{\infty}\varepsilon_0\!\int_V d{\bf x} \Big(\!\bnabla \left(\phi({\bf x}) 
- \alpha(\bnabla\cdot {\bf P}({\bf x})\right)\!\Big)^2\!- \int_V d{\bf x}~ {\cal \bf P}({\bf x}) \bcalE({\bf x}) - V[\bcalE({\bf x}), \phi({\bf x})].\nonumber 
\label{hamilt0a}
\end{eqnarray}
\\
The free energy is now cast into a completely local form and the corresponding form of the Euler-Lagrange equations are obtained straightforwardly. The transformation of the general non-local free energy Eq. (\ref{maineq1}) into a local form, {\sl i.e.}, containing only local kernels, is important for our later analysis of the boundary conditions.  

In deriving the Euler-Lagrange equations for the above free energy one needs to remember that for the free energy density dependent on the first derivatives of a field, $f_V(u, \bnabla u)$, the Euler-Lagrange equations have the standard form 
\\
\begin{eqnarray}
\frac{\partial f_V}{\partial u} - \bnabla \Big( \frac{\partial f_V}{\partial \bnabla u}\Big) =0, 
\label{modelEL1}
\end{eqnarray}
\\
while for the free energy density dependent on the first and second derivatives of a field, $f_V(u, \bnabla u, \nabla^2 u)$, the Euler-Lagrange equations then read as
\\
\begin{eqnarray}
\frac{\partial f_V}{\partial u} - \bnabla \Big( \frac{\partial f_V}{\partial \bnabla u}\Big) + \nabla^2 \Big( \frac{\partial f_V}{\partial \bnabla^2 u}\Big) = 0.
\label{modelEL2}
\end{eqnarray}
\\
With this in mind, the variation of the free energy with respect to polarization, $\delta {\bf P}({\bf x})$, leads to the following equation for the polarization vector 
\\
\begin{eqnarray}
&& u_P(0) \left[{\cal \bf P}({\bf x}) - \xi^2 \bnabla \left(\bnabla\cdot {\cal \bf P}({\bf x})\right) + \zeta^4 \nabla^2 \left(\bnabla\cdot {\cal \bf P}({\bf x})\right)\right] + \nonumber\\
&&~~~~~ + \alpha\varepsilon_\infty\varepsilon_0 \nabla^2 \Big( \phi({\bf x}) - \alpha \left(\bnabla \cdot{{\cal \bf P}}({\bf x})\right)\Big)  - \bcalE({\bf x}) = 0. 
\label{pol11}
\end{eqnarray}
\\
From the above equation it follows straightforwardly that the auxiliary field $\bcalE$ is coupled exclusively with non-electrostatic dipolar and hydration interactions.

The variation with respect to the auxiliary polarization field, $\delta \bcalE({\bf x})$, leads to the modified non-linear Langevin-Poisson constitutive relation 
\\
\begin{eqnarray}
 - {\cal \bf P}({\bf x}) {- \frac{\partial }{\partial  \bcalE({\bf x})} \upsilon\left(\bcalE({\bf x}), \phi({\bf x})\right)} = 0, 
\label{polar}
\end{eqnarray}
\\
while the variation with respect to the auxiliary electrostatic potential, $\delta \phi({\bf x})$, yields the modified non-linear Poisson-Boltzmann equation
\\
\begin{eqnarray}
\hspace*{-2.5cm} - \varepsilon_\infty\varepsilon_0 \nabla^2 \Big( \phi({\bf x}) - \alpha \left(\bnabla \cdot{{\cal \bf P}}({\bf x})\right)\Big) - 
{\left( \frac{\partial }{\partial \phi({\bf x})}  - \bnabla \left(\frac{\partial }{\partial \bnabla\phi({\bf x})}\right) \right) \upsilon\left(\bcalE({\bf x}), \phi({\bf x})\right)}  = 0. \nonumber \\
\label{pol22}
\end{eqnarray}
The three Euler-Lagrange equations, Eqs. (\ref{pol11}), (\ref{polar}) and (\ref{pol22}), correspond to the definition of the non-electrostatic part of the polarization auxiliary field, 
the polarization field constitutive equation, and the generalized Poisson-Boltzmann equation, respectively. 

These are the final equations valid in the bulk of the system. We next formulate and analyze the boundary conditions.

\subsection{\label{sec:level2}Boundary terms and boundary conditions}

In an inhomogeneous system, which is the standard situation in (bio-)colloid science, the bulk is delimited by boundary surfaces. Often the surface is idealized to the extent that it presumably imposes some kind of {\sl ad hoc} boundary conditions on the inhomogeneous mean-field solution. These types of - otherwise easily implementable approaches - are obviously insufficient to describe the intricacies of the solution-interface interactions. In fact boundary surfaces and their interactions with the vicinal solution ions as well as solvent molecules give no hint of universality, and a plethora of models exist to describe these surface-solution interactions. 
 
The existence of surfaces implies boundary conditions for the order parameters and auxiliary fields that {\sl have to be consistent} with the underlying free energies \cite{Podgornik2018}. In order to derive these boundary conditions we need to include the variation of the bulk fields as well as the surface fields into the Euler-Lagrange equations \cite{Avni2019}. For a free energy that decouples into a volume part with density $f_V(u, \bnabla u)$ and a surface part with density $f_S(u_S)$, index $S$ denoting the surface value,  the complete Euler-Lagrange equations Eq. (\ref{modelEL1}) then also contain a surface part
\\
\begin{eqnarray}
{\bf n}\cdot\left( \frac{\partial f_V}{\partial \bnabla u}\right)_S + \frac{\partial f_S}{\partial u_S} = 0 ~~~~~
\label{modelEL1S}
\end{eqnarray}
\\
where $\bf n$ is the boundary surface normal. For the field which enters the free energy also with a second derivative, $f_V(u, \bnabla u, \nabla^2 u)$, the surface part of the  Euler-Lagrange equations Eq. (\ref{modelEL2}) then reads as
\\
\begin{eqnarray}
{\bf n}\cdot \left(\big( \frac{\partial f_V}{\partial \bnabla u}\big) - \bnabla \big( \frac{\partial f_V}{\partial \bnabla^2 u} \big)\right)_S + \frac{\partial f_S}{\partial u_S} = 0.  
\end{eqnarray}
\\
In most of the existing literature one usually considers only the volume part because the surface-specific interactions are assumed to be absent. Nevertheless, the part played by the surface specific interactions has been recognized by several authors \cite{Cevc1982,Kanduc2014,Monet2021}.

This decoupling of the Euler-Lagrange equations is of course only possible if the free energy can be separated into the volume and surface parts and if it can be written as a purely local functional of the fields, which means that in Eq. (\ref{maineq1}) we need to integrate out the Coulomb non-local interactions leading to a local inverse Coulomb operator, which then finally yields a purely local functional given by Eq. (\ref{hamilt0a}). 

Decoupling the local free energy functional into a volume and surface part allows us to derive not only the bulk Euler-Lagrange equations, but also their surface counterpart, see Ref. \cite{Podgornik2021}, {\sl a.k.a.} the boundary conditions. We therefore start with the expression
\\
\begin{eqnarray}
\beta {\cal F} &=& \beta {\cal F}_{V}   +  \oint_S d^2{\bf x}~f_{S}(P_S, {\cal E}_S, \phi_S),
\end{eqnarray}
\\
with $\beta {\cal F}_{V}$ given by Eq. (\ref{hamilt0a}). We
assume that the order parameters and the auxiliary fields have independent surface variations $\delta\phi_S$, $\delta {\cal E}_S = \delta (\bcalE \cdot {\bf n})_S$ and $\delta P_S = \delta ({\bf P}\cdot{\bf n})_S$, where the index $S$ refers to the surface values of the variables, we then end up with the following boundary equations: the surface variation of the electrostatic potential, $\delta\phi_S$, yields
\\
\begin{eqnarray}
{\bf n}\cdot\left( - \varepsilon_\infty \varepsilon_0 \bnabla \left( \phi  - \alpha \left(\bnabla \cdot{{\cal \bf P}}\right)\right) - \frac{\partial \upsilon}{\partial \bnabla \phi}\right)_S + \frac{\partial f_S}{\partial \phi_S} = 0, ~~~
\label{BCES}
\end{eqnarray}
\\
where the subscript $S$ in the first term signifies that the subscripted bulk quantity needs to be taken at the surface. Clearly this equation generalizes the boundary condition for the PB equation with surface interactions \cite{Podgornik1989}; analogously the surface variation of the auxiliary field, $\delta \bcalE_S$, yields a single surface terms of the form
\\
\begin{eqnarray}
\frac{\partial f_S}{\partial {\cal E}_S} = 0,
\end{eqnarray}
\\
and finally the surface variation of the polarization field, $\delta P_S$, which can be deduced from the free energy Eq. (\ref{pol11}) and leads to
\\
\begin{eqnarray}
\hspace*{-2.5cm} u_P(0) \left[\xi^2 \left(\bnabla\cdot {\bf P}\right)_S  - \zeta^4 \nabla^2 \left(\bnabla\cdot {\bf P}\right)_S\right]  +
 \varepsilon_\infty \varepsilon_0 \alpha ~~{\bf n}\cdot \bnabla\Big( \phi  - \alpha \left(\bnabla \cdot{{\cal \bf P}}\right) \Big)_S + \frac{\partial f_S}{\partial P_S} = 0. \nonumber \\
\end{eqnarray}
The last two boundary conditions are specific to our approach and embody the fact that electrostatic and polarization fields represent separate and independent degrees of freedom in this system. 

The question now remains as to what is the surface free energy density that we refer to above. This issue has been detailed in Ref. \cite{Podgornik2018} within the mean-field approximation and remains unchanged in the present formulation.  Since the description of the surface is much less universal and much more model dependent than the equivalent description of the bulk, we cannot aspire to the same level of generality as for the bulk. In fact, we need to make certain additional assumptions at this point in order to proceed. 

We assume that there are surface specific polarization and charge interactions that contribute the analogous terms to the total surface free energy as in the volume case that we derived before. This leads us to the proposition
\\
\begin{eqnarray}
\hspace*{-2cm}&& f_{S}(P_S, {\cal E}_S, \phi_S) =  {\textstyle\frac12} u_S P_S^2 - P_S{\cal E}_S - \left( P_0 P_S  + \sigma_0 \phi_S\right)  + \nonumber \\
\hspace*{-1.5cm}&& + \frac{k_BT}{b^{2}} \ln{\Big( \lambda_D {\cal P}\big(\beta q_D s ~\vert \bcalE_S - (\bnabla \phi)_S\vert\big)  + \lambda_+ ~e^{(\beta e \phi_S - \alpha_+)} + \lambda_- ~e^{-(\beta e \phi_S + \alpha_-)} \Big) } \nonumber\\
\hspace*{-1.5cm} && =  {\textstyle\frac12} u_S P_S^2 - P_S{\cal E}_S - \left( P_0 P_S  + \sigma_0 \phi_S\right) + \upsilon_S({\cal E}_S, \phi_S), 
\label{Xi-11a}
\end{eqnarray}
\\
with ${\cal P}(u) = \sinh{u}/u$. 
Eq. (\ref{Xi-11a}) is of course just one, but a fairly general one, assumed model to describe the bounding surfaces. Polarizability terms could be included if one deems the form of the free energy is still not complicated enough. Here, $b$ is the size of the surface sites, which in general differs from the bulk lattice gas sites. The first two terms in the above expression quantify the surface specific, short-range polarization interactions, where $P_0$ is the surface density of polarization sources and $\sigma_0$ is the surface density of charge sources.

Inserting now this free energy into the boundary conditions we are left with the modified surface electrostatic boundary condition
\\
\begin{eqnarray}
{\bf n}\cdot\left( - \varepsilon_\infty \varepsilon_0 \bnabla \Big( \phi  - \alpha \left(\bnabla \cdot{{\cal \bf P}}\right)\Big) - \frac{\partial \upsilon}{\partial \bnabla \phi}\right)_S - \sigma_0 + \frac{\partial \upsilon_S}{\partial \phi_S} = 0, \nonumber\\
~
\label{BC-1}
\end{eqnarray}
(note the difference between $\upsilon$ and $\upsilon_S$), the modified surface constitutive relation
\\
\begin{eqnarray} \label{BC-2}
 - P_S + \frac{\partial \upsilon_S}{\partial {\cal E}_S} = 0,
\end{eqnarray}
\\
and the modified surface polarization boundary condition
\\
\begin{eqnarray}
&&u_P(0) \left[\xi^2 \left(\bnabla\cdot {\bf P}\right)_S  - \zeta^4 \nabla^2 \left(\bnabla\cdot {\bf P}\right)_S \right] + \nonumber\\
&& ~~~~~~~~  +  \varepsilon_\infty \varepsilon_0 \alpha ~{\bf n}\cdot\bnabla\Big( \phi  - \alpha \left(\bnabla \cdot{{\cal \bf P}}\right) \Big)_S + u_S P_S - {\cal E}_S - P_0 = 0. \nonumber\\
~
\label{BC-3}
\end{eqnarray}
Clearly the derived boundary conditions, consistent with the form of the bulk as well as the surface free energies, are nowhere close to the assumption of constant surface fields. The electrostatic field boundary condition, Eq. (\ref{BC-1}), is closely related to the boundary conditions used in the charge regulation theory \cite{Podgornik1986, Podgornik2021, Markovich2021}, while the polarization boundary condition, Eq. (\ref{BC-3}), is related to the boundary conditions used in the theory of hydration/structural forces \cite{Cevc1982, Abrashkin2007, Kanduc2014}.

Together with the bulk Euler-Lagrange equations these boundary conditions close the formulation of the non-homogeneous case. We next consider two illuminating limiting cases of the boundary condition obtained in the absence of hydration coupling, {\sl i.e.}, $\alpha = 0$, that reduce to more familiar forms.

\subsubsection{\label{B1} Limiting cases I: inner Helmholtz layer - surface polarization.}

The first interesting limiting case is obtained by ignoring any specific interactions of the ions with the surface, or actually ignoring ions altogether. In that case Eq. (\ref{Xi-11a}) simplifies to
\begin{equation}
\hspace*{-0.5cm} f_{S}(P_S, {\cal E}_S, \phi_S) = {\textstyle\frac12} u_S P_S^2 - P_S{\cal E}_S - P_0 P_S  + \frac{k_BT}{b^{2}} \ln{\Xi(\bcalE_S, (\bnabla \phi)_S)},
\label{Xi-11b}
\end{equation}
\vspace*{0.5cm} 
where now the surface partition function 
\\
\begin{eqnarray}
\Xi(\bcalE_S, (\bnabla \phi)_S) = 1 + \lambda_D {\cal P}(\beta q_D s \vert \bcalE_S - (\bnabla \phi)_S\vert)
\end{eqnarray}
\\
corresponds to the surface lattice gas of dipoles.  In this case the relevant boundary conditions are obtained as  
\\
\begin{eqnarray} \label{bc-inner1} 
- \varepsilon_\infty \varepsilon_0 ({\bf n}\cdot\bnabla \phi)_S + P_S = 0 \qquad {\rm with} \qquad P_S =  \frac{k_BT}{b^{2}}  \frac{\partial \ln{\Xi}}{\partial {\cal E}_S}, 
\label{BC-1a1}
\end{eqnarray}
\\
for the surface constitutive relation connecting the electrostatic field and the polarization field, as well as 
\\
\begin{equation} \label{bc-inner2}
u_P(0) \left[\xi^2 \left(\bnabla\cdot {\bf P}\right)_S  -  \zeta^4 \nabla^2 \left(\bnabla\cdot {\bf P}\right)_S\right] + u_S P_S - {\cal E}_S - P_0 = 0.
\end{equation}
\vspace*{0.5cm} 
The solution of these boundary conditions gives us the surface polarization at the inner Helmholtz plane as a function of the surface electrostatic field. 
Because of the logarithm appearing in Eq. (\ref{BC-1a1}), the surface polarization reads as
\\
\begin{eqnarray} \label{bc-inner3} 
P_S =  \frac{\lambda_D q_D s}{b^{2}}  \frac{\frac{\partial {\cal P}(u)}{u ~\partial {u}} (\bcalE_S - (\bnabla \phi)_S)}{1 +  {\cal P}(\beta q_D s \vert \bcalE_S - (\bnabla \phi)_S\vert)}, 
\label{BC-1a1a}
\end{eqnarray}
\\
and clearly shows a saturation behavior that cannot exceed the value of a fully oriented layer of surface solvent molecules. 

\subsubsection{Limiting cases II: outer Helmholtz layer - charge regulation.}

We obtain another interesting limit of the above general model by first assuming that there are no polarization effects at the surface. In that case Eq. (\ref{Xi-11a}) becomes
\\
\begin{eqnarray}
&& f_{S}(P_S, {\cal E}_S, \phi_S) = - \sigma_0 \phi_S +\frac{k_BT}{b^{2}} \ln{\Xi_S(\phi_S)}, 
\label{Xi-11b2}
\end{eqnarray}
\\
with the surface partition function 
\\
\begin{eqnarray}
\Xi_S(\phi_S) =  1 + \lambda_+ ~e^{(\beta e \phi_S - \alpha_+)} + \lambda_- ~e^{-(\beta e \phi_S + \alpha_-)}
\end{eqnarray}
\\
that obviously corresponds to a surface lattice gas of adsorbed ions, or equivalently to a Langmuir isotherm. 
Of the different boundary conditions only Eq. (\ref{BC-1}) remains relevant and it can be recast in the form
\\
\begin{eqnarray}
- \varepsilon_\infty \varepsilon_0 ({\bf n}\cdot\bnabla \phi)_S + {P}_S = \sigma_0 - \sigma(\phi_S),
\label{BC-1a}
\end{eqnarray}
\\
where we introduced
\\
\begin{equation}
\sigma(\phi_S) = \frac{\partial \upsilon_S}{\partial \phi_S} = \frac{\lambda_+\beta e}{\Xi_S(\phi_S)} ~e^{(\beta e \phi_S - \alpha_+)} - \frac{\lambda_-\beta e}{\Xi_S(\phi_S)} ~e^{-(\beta e \phi_S + \alpha_-)}.
\end{equation}
\vspace*{0.5cm} 
In the limit of adsorption for only a single type of ion, {\sl e.g.} $\alpha_- \longrightarrow \infty$, we get the boundary condition Eq. (\ref{BC-1a}) in the simplified form 
\\
\begin{eqnarray} \label{bc-outer}
\varepsilon_\infty \varepsilon_0 E_S + P_S = \sigma_0 - 2 \lambda_+\beta e \Big( 1 + \tanh{{\textstyle\frac12 } (\beta e \phi_S - \alpha_+)}\Big), ~~~
\end{eqnarray}
\\
corresponding exactly to the Langmuir adsorption isotherm charge regulation \cite{Ninham1971,Podgornik2021}. The {\sl Ansatz} for the surface free energy, Eq. (\ref{Xi-11a}), thus in general also incorporates charge regulation. 

\section{\label{sec:solution}Solving the model equations}

\subsection{Generalities} 

In our previous work we have discussed different versions of the bulk model in order to clarify the role the different bulk lengths play in the polarization interactions \cite{Blossey2022}. 
Specifically, we solved the linearized equations in a one-dimensional geometry with analytical and numerical methods.

The inclusion of further physical effects and, in particular, in the newly formulated boundary condition in this work renders the resulting theory extremely rich in physical variables and parameters. Not all of them will be of equal relevance for a specific experimental system, such that the determination of completely general parameter diagrams from the present theory will hardly be of general value. 
In this section we therefore point out general characteristics of our theory which will be useful in later applications. As in our previous work we consider the implementation of our theory for a 1D case, corresponding to a single or two planar surfaces perpendicular to the axis $z$, and in the latter case separated by the distance $L$.
In 1D the spatial dependence is a single coordinate $z$ and all the vectorial variables only retain their $z$ component, {\sl i.e.}, 
\\
\begin{eqnarray}
{\bf P} = (0,0, P(z)),  ~\bnabla \phi = (0,0, \phi'(z)),  ~\bcalE = (0,0, {\cal E}(z)).~~~
\end{eqnarray}
\\
We can write Eq. (\ref{hamilt0a}) as
\\
\begin{eqnarray}
\beta {\cal F}_V && \equiv    S~ \int_L d{z}~ f\left({P}, ~{P}', ~{P}''; ~\phi,  ~\phi'; ~\bcalE \right), 
\end{eqnarray}
\\
where $S$ is the surface area. A useful general result is now available in the form of a first integral, or the stress tensor for a one dimensional system \cite{Budkov2022}. In this case it can be obtained following the approach described in Ref. \cite{Gruen1983-2} as the expression
\\
\begin{eqnarray} \label{firstintegral} 
f - \phi'\frac{\partial f}{\partial \phi'} - P'\frac{\partial f}{\partial P'} - P''\frac{\partial f}{\partial P''}  + P' \left( \frac{\partial f}{\partial P''}\right)' = const. \, ~~~~
\end{eqnarray}
\\
The two last terms are a consequence of the higher-order derivative terms in the free energy density, see our previous discussion of the Euler-Lagrange
equations in Sec. 2.  This relation is of general use in the integration of the saddle-point equations, albeit in the present complicated case this is not as easy to implement as
for the standard PB-equation for simple ions, where the electrostatic potential can be obtained by exact integration of the first integral. We stress that Eq. (\ref{firstintegral}) 
holds for the fully nonlinear equations. 

Of crucial interest is the role of the boundary conditions given for the general case by Eqs. (\ref{BC-1}), (\ref{BC-2}) and (\ref{BC-3}); we consider them for one surface only; we also set $\alpha = 0$. 
In the case of the two surfaces of the slit geometry, the discussion needs to be adapted accordingly. 

The first important element is to notice the interdependence of the different surface variables expressed in these equations. The theory contains the six surface variables 
\\
\[
{\cal E}_S,\, \phi_S,\, \phi_S',\, P_S,\, P_S',\, P_S''',
\]
\vspace*{0.5cm} 
whose relations are determined by the three boundary equations. Ultimately, this means that three values of the variables remain to be chosen. 

Looking more closely, Eq.(\ref{BC-1}) expresses a relation between the field values ${\cal E}_S$, 
$\phi_S $ and $\phi_S' $. We can thus, e.g., obtain ${\cal E}_S$ as a function of $\phi_S$ and $\phi_S' $, which together with Eq. (\ref{BC-2}) then defines a function $P_S = g_2(\phi_S, \phi_S')$. Fixing the values for ${\cal E}_S$ and $P_S$ then allows to determine the corresponding values of $\phi_S$ and $\phi_S'$. Further, Eq.(\ref{BC-3}) generally expresses a relationship between
$P_S$, ${\cal E}_S$, $\phi_S'$ and the higher-order derivatives of $P_S$. In fact, one now has 
\\
\begin{equation}
    u_P(0) (\xi^2 P_S''  - \zeta^4 P_S'''') = g_3({\cal E}_S, P_S,\phi_S')\, .
\end{equation}
\vspace*{0.5cm} 
From the previous expressions the right-hand side of the last equation is now entirely fixed through the solution of the two previous equations. Thus, this relation fixes the relative difference between the two higher-order derivatives which remains as a final choice of boundary conditions. In this exemplary construction, the three values to choose are thus $({\cal E}_S, P_S) $ and the difference of the higher derivatives.

We finally illustrate this reasoning for the two limiting cases of surface polarization and charge regulation. 
In the case of the first example, the inner Helmholtz later, the surface polarization case which ignores the ions, since the boundary conditions do not depend on the electrostatic potential $\phi$, Eq. (\ref{BC-1a1}) result in a functional relationship between
${\cal E}_S $ and $P_S$. In general, this equation can only be resolved numerically. If one expands both
${\cal P}$ and the $\ln$ for small arguments, ${\cal E}_S$ turns out to be directly proportional to $P_s$. 
In this limit Eq.(\ref{BC-3}), the relative difference in the polarization derivatives at the wall, becomes
a linear function of $P_s$, while the surface free energy becomes a quadratic function of $P_s$, as is 
well-known. We now consider our two limiting examples, the inner and outer Helmholtz layer cases, in detail.

\subsection{The outer and inner Helmholtz layers: examples in 1D} 

In order to illustrate the role played by the explicit boundary conditions, we first analyze the case of the outer Helmholtz layer for the linearized theory in the case of $\zeta =0 $. 
For the bulk part of the system this corresponds to our Model 1 of \cite{Blossey2022}. 

We begin the discussion by recapitulating the linearized model equations in the bulk. As in Refs. \cite{Gruen1983,Paillusson2010} we introduce the {\sl polarization potential}
$\phi^{\dagger}({\bf x})$ which can be defined via 
\begin{equation}
{\bf P}({\bf x}) = (\varepsilon - \varepsilon_{\infty})\varepsilon_0 \bnabla\phi^{\dagger}({\bf x})\, ,
\label{polfield}
\end{equation}
\vspace*{0.5cm} 
making use of the identification 
\\
\begin{equation}
     \varepsilon \varepsilon_0 \equiv \varepsilon_{\infty}\varepsilon_0 + \frac{{\textstyle\frac{1}{3}} \lambda p^2}{1 + u_P(0){\textstyle\frac{1}{3}} \lambda p^2} 
\label{def-epsilon2}     
\end{equation}
\vspace*{0.5cm} 
that relates the structural coupling strength $u_P(0)$ to the dielectric constants and the strength of the water dipole. With the definition Eq. (\ref{def-epsilon2}) and, additionally, setting
\\
\begin{equation}
\widehat{\xi}^2 \equiv (\varepsilon - \varepsilon_{\infty})\varepsilon_0 u_P(0)\, \xi^2, 
\end{equation}
\vspace*{0.5cm} 
the mean-field equations for our Model 1 read as, using the notation of \cite{Blossey2022},
\\
\begin{eqnarray}
\bnabla^2\phi^{\dagger}({\bf x}) &=& {\widehat{\xi}^{-2}} \left( \phi^{\dagger}({\bf x}) - \phi^*({\bf x})\right),\\
 \bnabla^2\phi^*({\bf x})  &=& ~ {\textstyle\frac{\varepsilon}{\varepsilon_{\infty}}} \kappa_D^2 \phi^*({\bf x})  
 +\,\, {\widehat{\xi}^{-2}}\left( {\textstyle\frac{\varepsilon}{\varepsilon_{\infty}}} - 1 \right) \left( \phi^*({\bf x}) - \phi^{\dagger}({\bf x})\right)\,, \nonumber
\label{sp3c}
\end{eqnarray}
\\
in which the inverse square of Debye length is defined by $\kappa_D^2 \equiv {2(\beta e)^2 \lambda_s}/{\varepsilon\varepsilon_0}$. Except for the sign of $\phi^{\dagger}$ these equations are the same as the Onsager-Dupuis equations \cite{Onsager1960}. 

The model equations can be written in matrix form $\frac{d^2}{dz^2} \Phi_1(z) = {\cal M}_1 \Phi_1(z)$ introducing the composite field $\Phi_1(z) \equiv (\phi^*(z),\phi^{\dagger}(z))$. 
The matrix ${\cal M}_1$ is given by
\\
\begin{eqnarray}
\hspace*{-0.5cm}{\cal M}_1 = \left(\begin{array}{cc}
    \frac{\varepsilon}{\varepsilon_{\infty}}\kappa_D^2 + \left(\frac{\varepsilon}{\varepsilon_{\infty}} - 1\right)\widehat{\xi}^{-2} &  - \left(\frac{\varepsilon}{\varepsilon_{\infty}}- 1\right)\widehat{\xi}^{-2} \\
    -\widehat{\xi}^{-2}      &  \widehat{\xi}^{-2} 
\end{array}\right)\, . \nonumber
\end{eqnarray}
\\
The diagonalization of the matrix and a rescaling of its eigenvalues with $\lambda \longrightarrow \lambda/\widehat{\xi}^2$ results in a quadratic eigenvalue equation given by
\\
\begin{equation}
\lambda^2 - \lambda~ {\textstyle\frac{\varepsilon}{\varepsilon_{\infty}}}\left( 1 + (\kappa_D\widehat{\xi})^2\right) + {\textstyle\frac{\varepsilon}{\varepsilon_{\infty}}}(\kappa_D\widehat{\xi})^2 = 0,
\end{equation}
\vspace*{0.5cm} 
where $\kappa_{1,2} = \sqrt{\lambda_{1,2}}$ are the two inverse decay lengths corresponding to the two eigenvalues, which are both real and positive. They correspond to the decay lengths of the electrostatic and the polarization potentials, respectively, and coincide with the result of the linearized Onsager-Dupuis theory \cite{Gruen1983,Paillusson2010}. 

The Model 1-equations have been solved analytically before, e.g. for the slit geometry explicitly in \cite{Paillusson2010}, and numerically by us in
\cite{Blossey2022}. We do not wish to recapitulate these results here, but would like to focus on what solutions are selected by the choice of the new boundary conditions. 
This can most easily be done already for the case of a single plate, so that we can focus on only one set of boundary conditions. Thus we have the solutions
\\
\begin{eqnarray} \label{ansatz}
\phi^*(x) & = & A e^{-\kappa_1 x} + B e^{-\kappa_2 x}\\
\phi^{\dagger}(x) & = & C e^{-\kappa_1x} + D e^{-\kappa_2 x}\, , 
\end{eqnarray}
\\

\begin{figure}[th] 
\begin{center}
{\bf a)} \includegraphics[width=10cm]{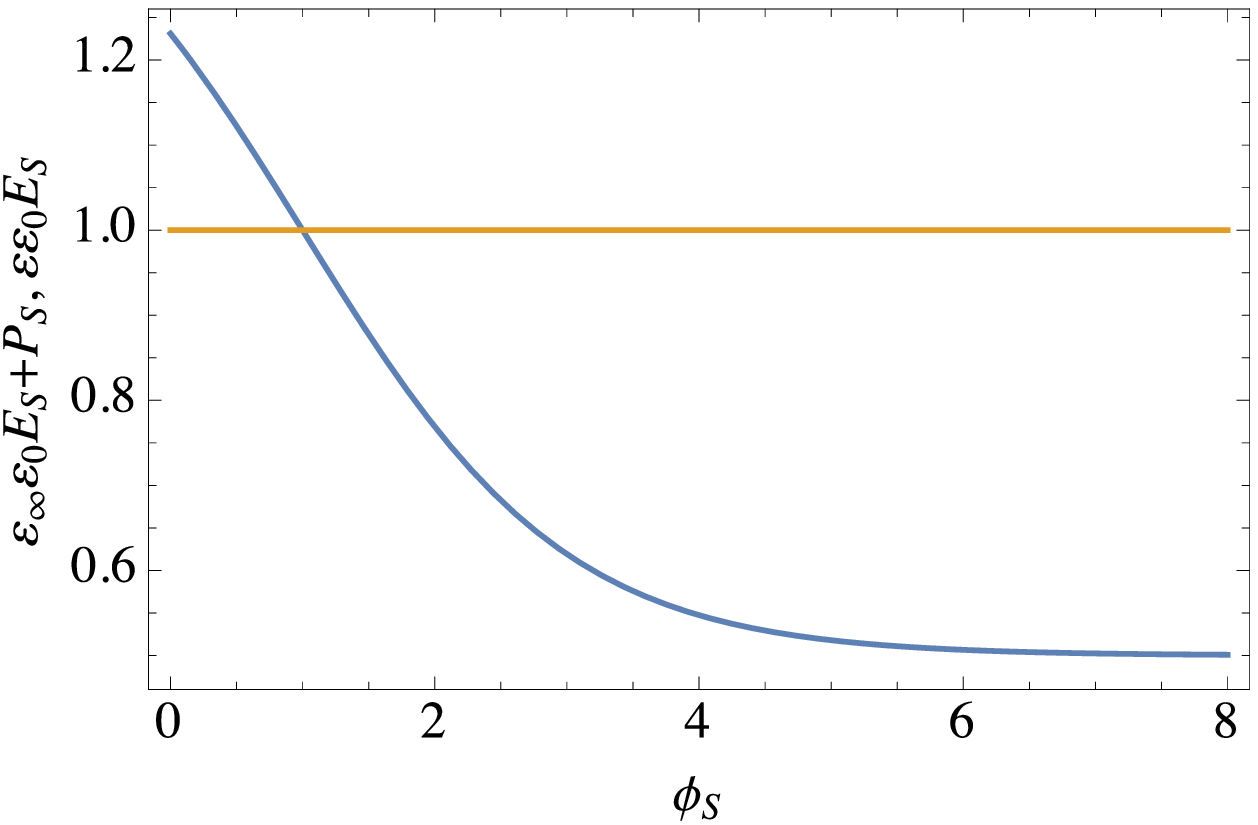}
\\
{\bf b)} \includegraphics[width=10cm]{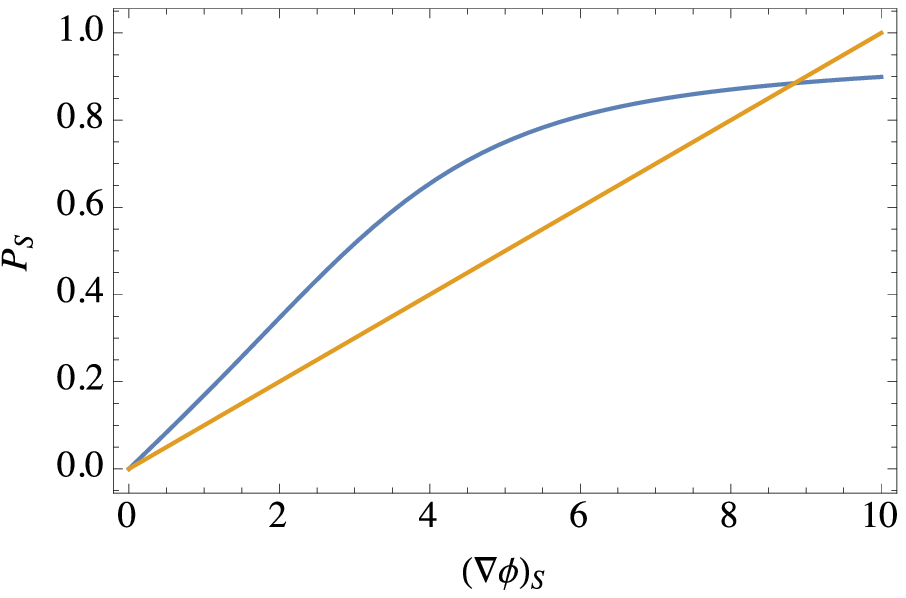}
\end{center}
\caption{a) the nonlinear boundary condition, Eq. (\ref{bc-outer}). Blue curve: $\varepsilon_{\infty}\varepsilon_0 E_S + P_S $; yellow straight line $ \varepsilon \varepsilon_0 E_S$; 
b) The graphical solution to Eqs. (\ref{bc-inner1}), (\ref{bc-inner3}).}     
\label{Figure}
\end{figure} 
which when inserted in the equations yields the system
\\
\begin{eqnarray} \label{bulk}
\kappa_1^2 C + \kappa_2^2 D = \widehat{\xi}^{-2}\Big((C+D) - (A+B)\Big)\\
\kappa_1^2 A + \kappa_2^2 B = \frac{\varepsilon}{\varepsilon_{\infty}}\kappa_D^2 (A+B) + \widehat{\xi}^{-2}\Big((A+B) - (C+D)\Big)
\end{eqnarray}
\\
which makes up for two equations in four unknowns.  The two additional equations for the coefficients are provided by the boundary conditions, as in \cite{Blossey2022}:
\\
\begin{eqnarray} \label{surface} 
E_S \sim {\phi^*}'(0) = - \kappa_1 A - \kappa_2 B\\
P_S \sim {{\phi}^{\dagger}}'(0) = - \kappa_1 C - \kappa_2 D\,.
\end{eqnarray}
\\
The value of $\phi_S$ at the surface is given by
\\
\begin{equation} \label{surface-phi}
\phi_S \sim \phi^*(0) = A + B\,.  
\end{equation}
\vspace*{0.5cm} 
Thus, the two bulk equations Eqs. (\ref{bulk}), coupled with the two boundary conditions, Eqs. (\ref{surface}), allow to determine
the four amplitudes $A-D$ in Eq.(\ref{ansatz}) and thus to completely solve the problem. Eq. (\ref{surface-phi}) then follows as a consequence:
the value of the electrostatic potential is fixed by the bulk equations and the choice of the gradients of the electrostatic and polarization potential at the surface. 

This now is the crucial point where the new boundary conditions change matters considerably. Eq. (\ref{bc-outer}) couples $E_S, P_S $ and $\phi_S$, and this follows for the amplitudes $A-D$ when inserting the conditions (\ref{surface}) and (\ref{surface-phi}) into this equation:

\begin{equation}
\varepsilon_\infty \varepsilon_0 E_S + P_S = \sigma_0 - 2 \lambda_+\beta e \Big( 1 + \tanh{{\textstyle\frac12 } (\beta e \phi_S - \alpha_+)}\Big)\, .
\end{equation} 
\vspace*{0.5cm}
Simplifying by linearizing the $\tanh$ we have
\\
\begin{equation}
\varepsilon_\infty \varepsilon_0 E_S + P_S = \sigma_0 - 2 \lambda_+\beta e \Big(1 + \frac{1}{2} \beta e \phi_S - \alpha_+\Big)
\end{equation}
\vspace*{0.5cm}
which one can now rewrite in terms of the amplitudes $A-D$ of the bulk solution:
\\
\begin{equation}
\hspace*{-1.5cm} -\varepsilon_\infty \varepsilon_0 (\kappa_1 A + \kappa_2 B) - (\kappa_1 C + \kappa_2 D) = 
\sigma_0 - 2 \lambda_+\beta e \Big(1 + \frac{1}{2} (\beta e (A + B) - \alpha_+)\Big)\, .
\end{equation}
\vspace*{0.5cm} 
In this case there are only three equations in terms of the amplitudes $A-D$, and in general, the relation between the coefficients is in fact even a nonlinear one, as is sketched in Figure 1 a). The gradients of the electrostatic and polarization potential at the surface are now turned into explicit nonlinear functions of the value of the electrostatic potential at the surface.

The same reasoning applies to the problem of the outer Helmholtz layer.
In addition to the two bulk equations, Eqs. (\ref{sp3c}) 
one now has to find a 
self-consistent solution to Eqs.(\ref{bc-inner1}) and (\ref{bc-inner3}). A graphical solution is shown in Figure 1 b) for the case ${\cal E}_S = 0$. 
This solution then serves as input to Eq. (\ref{bc-inner2}) which, in our simplified case reads as
\\
\begin{equation}
- u_P(0) \xi^2 \nabla^2 \phi^{\dagger}_S + u_S P_S = P_0\, .      
\end{equation}
\vspace*{0.5cm} 
In the case ${\cal E}_S \neq 0$, Eqs. (\ref{bc-inner1}) has to be
numerically solved for ${\cal E}_S$, with the solution being the
input to Eq. (\ref{bc-inner2}). Solving these equations
then yields the remaining coefficients.

\section{\label{sec:discussion}Conclusions and Outlook}

In this work we have derived a comprehensive continuum theory of structured inhomogeneous liquid dielectrics. 
Starting from previous work for the bulk case \cite{Blossey2022}, we develop a substantially extended version
of our theory which explicitly includes the polarizability of the solvent molecules and the solvated ion species.
In particular, we have carefully developed the theory of the surface free energy which includes surface
polarization and charge regulation as two generic mechanisms. After formulating the bulk and surface free energies,
we showed how to derive the corresponding saddle-point equations (nonlinear mean-field equations) together
with the boundary equations. We discussed the role the boundary conditions play in the selection of the solutions by 
regarding two explicit examples of an inner and outer Helmholtz layer. 

It is useful to put our model in the context of earlier models for structural interactions in liquid dielectrics and results obtained from
them. As we have discussed in detail in our review \cite{Blossey2022-2}, 
the class of models based entirely on the polarization field has so far largely focused on the ion-free case or single solvated ions. On the other hand, there are the dipolar Poisson-Boltzmann theories built on the 
single order parameter of the electrostatic potential \cite{Abrashkin2007}.
Both these limiting theories and the results obtained from them are firmly contained in our theory. Combined theories,
such as the Onsager-Dupuis theory itself, which is again contained in
our theory, have so far never been systematically studied. This is, in 
particular, the case for the boundary conditions of such theories formulated
in the two fields of the electrostatic field $\phi$ and the polarization
field ${\bf P}$. This significant extension will allow the application of our theory to real liquid dielectric systems. Our present study thus opens a new way to  treat electrostatic and structural forces in liquid dielectrics systematically and on the same footing.
Being a fully formulated field theory, its application beyond the mean-field approximation can be achieved by a computation of corrections in terms of a loop expansion. 

We therefore believe that our novel formulation of a comprehensive continuum theory of structured liquid dielectrics
meets the challenge raised by Mugele et al. that we cited in the Introduction - it seems to be well possible
to formulate a general Poisson-Boltzmann theory containing all non-electrostatic effects. We are careful to
rather state `general' than `universal' here, since many of the included physical effects are indeed highly
specific and thus a properly `universal' theory cannot exist in principle. 

The next step in the formulation of the theory will be the confrontation with experiment. Clearly, our comprehensive theory contains special cases for which such comparisons with experiment have already
been undertaken; the challenge therefore rather is: how far can we go? As already discussed in the Introduction
we think that the application to AFM measurements of structural forces are an evident first step to take.
Previous work by Benaglia et al. \cite{Benaglia2021} went already along this path employing density-functional theory (DFT). We think that the main advantage of the field-theoretic approach is
its high transparency of the physical mechanisms explicitly built into the theory, in particular the flexibility of the treatment of the boundary conditions. This point of view is strengthened by the recent AFM-experiments 
of ionic solutions on silica and gibbsite surfaces \cite{Klaassen2022}. The experimentally observed oscillations in the force-distance curves are clearly due the ordering of bulk water molecules in the vicinity of the surface. The details of these curves are nevertheless dependent on pH and hence on charge regulation effects. We believe
that our approach will be flexible enough to disentangle and identify these different relevant effects on
electrostatic and hydration forces at complex substrate surfaces.
\\

{\bf Acknowledgments.} RP wishes to acknowledge the support of the University of Chinese Academy of Sciences and funding from the NSFC under Grant No. 12034019.
RB thanks F. Mugele, I. Siretanu and S. Kumar for discussions on the application of the present theory to experiments.
\\

{\bf Data availability statement.} All data that support the findings of this study are included within the article (and any supplementary files).
\\

{\bf Appendices}
\\

{\bf A. Drude model}
\\

The Drude model of polar and polarizable molecules \cite{Bordin2016} is based on two oppositely charged 
particles connected by a harmonic spring potential \cite{Buyukdagli2013,Budkov2020}
\\
\begin{eqnarray}
U(\bell , {\bf n}) = {\textstyle\frac12}\frac{(\bell - {\bf s})^2}{ \sigma^2} = {\textstyle\frac12}\frac{(\bell - s{\bf n})^2}{\sigma^2},
\end{eqnarray}
\\
where ${\bf n} = {\bf s}/\vert {\bf s}\vert = {\bf s}/s$, yielding the separation distribution function 
$g(\ell)$ in the form
\\
\begin{equation}
\hspace*{-1.5cm} g(\bell) = (2\pi \sigma^2)^{-3/2} \int_{\Omega} \frac{d{\bf n}}{4\pi} ~\exp{- \frac{(\bell - {\bf s})^2}{2 \sigma^2}} = (2\pi \sigma^2)^{-3/2} \int_{\Omega} \frac{d{\bf n}}{4\pi} ~e^{-U(\bell, {\bf n} )}
\label{byfcuw}
\end{equation}
\vspace*{0.5cm} 
with an average separation $\bf s$ and its variance $\sigma^2$ given by 
\\
\begin{eqnarray}
\mathopen< \bell \mathclose> = s, \qquad \mathopen< (\bell - {\bf s})^2 \mathclose> = \sigma^2,
\end{eqnarray}
\\
where the average was defined as
\\
\begin{eqnarray}
\mathopen< \dots \mathclose> &=& \int d\ell g(\ell) = (2\pi \sigma^2)^{-3/2}  \int d\ell \int_{\Omega} \frac{d{\bf n}}{4\pi} (\dots ) ~e^{-U(\bell, {\bf n} )}.~~~~~
\label{harmonic}
\end{eqnarray}
\\
This is the model that we use to represent the aqueous solvent molecules as well as the solvated ions. The Drude model is valid for small separation between the particles where the difference in the electrostatic potential felt by each of them can be approximated to the lowest order in the Taylor expansion by the gradient in the potential.
\\

{\bf B. One-body partition function}
\setcounter{equation}{0}
\\

Here we derive the two expressions for the one-particle partition functions, $V[\bcalE_i({\bf x}), \phi({\bf x}))]$,  in external fields which feature in the filed-theoretic description of our system. The derivations follow Ref. \cite{Abrashkin2007} except for the inclusion of the polarizability terms. In addition, at the end because we work exclusively on the saddle-point level, we will need to make the transform  $\phi \longrightarrow i \phi$,  $\bcalE \longrightarrow i \bcalE$. 

The one-body partition function in an external field, $V[\bcalE_i({\bf x}), \phi({\bf x}))]$, is  defined as 
\\
\begin{equation}  \label{1} 
e^{V[\bcalE_i({\bf x}), \phi({\bf x}))]} \equiv \sum_{N}\ \sum_{N^+}{\sum_{N^-}} \frac{\lambda^N}{N!}  \frac{\lambda_{(+)}^{N^+} \lambda_{(-)}^{N^-}}{N^+!N^-!} 
\times  ({\cal I}_N {\cal I}_{N^+} {\cal I}_{N_-})
\end{equation}
where 
\\
\begin{eqnarray} 
{\cal I}_N \equiv \int {\cal D}[{\bf x}_N]~ {\cal D}[\ell_N] {\cal D}[{\bf n}_N] ~e^{- \sum_N U(\ell_i, {\bf n}_i)} 
\times  \\
~e^{-i \beta q_D \sum_N (\phi({\bf x}_i) - \phi({\bf x}_i + \ell_i)) - i~ q_D   \sum_N {\ell}_i ~\! \bcalE({\bf x}_i)} \nonumber 
\end{eqnarray} 
\begin{eqnarray}
{\cal I}_{N^+} \equiv \int {\cal D}[{\bf x}_+]{\cal D}[\ell_+] e^{- \sum_{N^+} U(\ell_i, {\bf n}_i)} 
\times  \\
~e^{ -i \beta \sum_{N^+}\phi({\bf x}_i) -i \beta q_I \sum_{N^+} (\phi({\bf x}_i) - \phi({\bf x}_i + \ell_i))- i~ q_I   \sum_{N^+} {\ell}_i ~\! \bcalE({\bf x}_i)} \nonumber 
\end{eqnarray}
\begin{eqnarray} 
{\cal I}_{N^-}  \equiv \int {\cal D}[{\bf x}_-]{\cal D}[\ell_-] e^{- \sum_{N^-} U(\ell_i, {\bf n}_i)} 
\times  \\
~e^{ i \beta \sum_{N^-}\phi({\bf x}_i) -i \beta q_I \sum_{N^+} (\phi({\bf x}_i) - \phi({\bf x}_i + \ell_i))- i~ q_I   \sum_{N^-} {\ell}_i ~\! \bcalE({\ell}_i)}. \nonumber 
\label{Xi-00}
\end{eqnarray}
\\

Using the definition of the average over the harmonic Drude degrees of freedom, Eq. (\ref{harmonic}), for polar solvent molecules (with $\mathopen< \bell \mathclose> = {\bf s}s$ and $\mathopen< (\bell - {\bf s})^2 \mathclose> = \sigma_D^2$) and ionic charges (with $\mathopen< \bell \mathclose> = 0$ and $\mathopen< \ell^2 \mathclose> = \sigma_I^2$) we end up with
\\
\begin{eqnarray}
\hspace*{-1.5cm} \int {\cal D}[\bell_N] {\cal D}[{\bf n}_N]~e^{- \sum_N U(\bell_i, {\bf n}_i)} e^{-i \beta q_D \sum_N (\phi({\bf x}_i) - \phi({\bf x}_i + \bell_i))- i~ q_D~\sum_N{\bell}_i\cdot \bcalE({\bf x}_i)} = \nonumber\\
= \Big(\frac{\sin{\left(\beta q_D s~ \vert\bnabla\phi({\bf x}) -  \bcalE({\bf x})\vert\right)}}{\beta q_D { s}~\vert\bnabla\phi({\bf x}) -  \bcalE({\bf x})\vert } ~e^{-{\textstyle\frac12}{ (\beta\sigma q_D)^2}\left(\bnabla\phi({\bf x}) -  \bcalE({\bf x})\right)^2}\Big)^N \nonumber
\end{eqnarray}
\\
\begin{eqnarray} 
\hspace*{-1.5cm}\int {\cal D}[{\bf x}_+] ~e^{ -i \beta \sum_{N^+}\phi({\bf x}_i)} &=& \Big(\int d^3{\bf x} ~e^{ -i \beta \phi({\bf x})}~e^{-{\textstyle\frac12}{ (\beta\sigma q_I)^2}\left(\bnabla\phi({\bf x}) -  \bcalE({\bf x})\right)^2}\Big)^{N^+} \nonumber\\
\hspace*{-1.5cm}\int {\cal D}[{\bf x}_-] ~ e^{ i \beta \sum_{N^-}\phi({\bf x}_i)} &=& \Big(\int d^3{\bf x} ~e^{ i \beta \phi({\bf x})}~e^{-{\textstyle\frac12}{ (\beta\sigma q_I)^2}\left(\bnabla\phi({\bf x}) -  \bcalE({\bf x})\right)^2}\Big)^{N^-}.
\end{eqnarray}
\\
Clearly all the sums in Eq. (\ref{Xi-00}) can be evaluated explicitly so that we finally remain with
\begin{eqnarray}
\hspace*{-1.5cm} V[\bcalE({\bf x}), \phi({\bf x})] = \\ 
\int_V d^3{\bf x} ~\Big( \lambda_D 
\frac{\sin{\left(\beta q_D s~ \vert\bnabla\phi({\bf x}) -  \bcalE({\bf x})\vert\right)}}{\beta s~\vert\bnabla\phi({\bf x})-  \bcalE({\bf x})\vert} ~ e^{-{\frac{(\beta\sigma q_D)^2}{2}}\left(\bnabla\phi({\bf x}) -  \bcalE({\bf x})\right)^2} \Big. + \nonumber \\
\hspace*{2cm} +  \Big. 2 \lambda_s  \cos{\beta e \phi({\bf x})}~e^{-{\frac{(\beta\sigma q_I)^2}{2}}\left(\bnabla\phi({\bf x}) -  \bcalE({\bf x})\right)^2}\Big). \nonumber \\
\label{bgwfxewa}
\end{eqnarray}

We now repeat the procedure for the case of the lattice gas. The only difference is in how one treats the sum over particles which now has to be performed over the discrete sites of the lattice.
Introducing the occupation of sites by electrolyte ions and Drude ions, which will be taken as occupying a single cell, we assign to each cell $j$, located at ${\bf x}_j$, a spin-like variable 
$s_j$ that can have one of three values: $s_j = 0$ if the cell is occupied by a Drude dipole, and $s_j = \pm 1$ according to the sign of the electrolyte ion.
Defining a new variable that depends on the spin-like variable $s_j$, $u(s_j)$, as
\\
\begin{eqnarray}
u(s_j) &=& i \beta e ~s_j \phi({\bf x}_j) +i \beta q_I s_j^2\left( \phi({\bf x}_j) - \phi({\bf x}_j + \bell_j)\right) \ \nonumber\\
&& - i \beta q_I s_j^2 ~\left(\bell_j\cdot \bcalE({\bf x}_j)\right) + \mu_j s_j^2 + \nonumber\\
&& +i \beta q_D (1-s_j^2)\left( \phi({\bf x}_j) - \phi({\bf x}_j + \bell_j)\right) - \nonumber\\
&& - i \beta q_D (1-s_j^2) ~\left(\bell_j\cdot \bcalE({\bf x}_j)\right) + \mu_j (1-s_j^2),
\end{eqnarray}
\\
it then follows specifically for $s_i = \pm1, 0$ that
\\
\begin{eqnarray}
u(s_j = +1) &=&  i \beta e ~ \phi({\bf x}_j) + i \beta q_I \left( \phi({\bf x}_j) - \phi({\bf x}_j + \bell_j)\right) + \nonumber\\
&& - i \beta q_I ~\left(\bell_j\cdot \bcalE({\bf x}_j)\right) + \mu_+ \nonumber\\
u(s_j = -1) &=&  -i \beta e ~ \phi({\bf x}_j) + i \beta q_I \left( \phi({\bf x}_j) - \phi({\bf x}_j + \bell_j)\right) + \nonumber\\
&& - i \beta q_I ~\left(\bell_j\cdot \bcalE({\bf x}_j)\right) + \mu_- \nonumber\\
u(s_j = 0) &=& i \beta q_D \left( \phi({\bf x}_j) - \phi({\bf x}_j + \bell_j)\right) - \nonumber\\
&& - i \beta q_D ~\left(\bell_j\cdot \bcalE({\bf x}_j)\right) +  \mu_D.  
\end{eqnarray}
\\
The one-particle partition function can then be written as
\\
\begin{equation}
\hspace*{-2cm} e^{V[\bcalE_i({\bf x}), \phi({\bf x}))]} \equiv \Pi_j \left( \sum_{s_j} e^{u(s_j)}\right) = 
\Pi_j \left( e^{u(s_j = +1)} + e^{u(s_j = -1)} + e^{u(s_j = 0)}\right).
\label{Xi-00a}
\end{equation}
\vspace*{0.5cm} 
From here it follows furthermore that
\\
\begin{eqnarray}
\hspace*{-2cm} V[\bcalE({\bf x}), \phi({\bf x})] = \sum_j 
\ln\Big(\lambda_D \Big.
\frac{\sin{\left(\beta q_D s~ \vert\bnabla\phi({\bf x}_j) - \bcalE({\bf x}_j)\vert\right)}}
{\beta s~\vert\bnabla\phi({\bf x}_j)- \bcalE({\bf x}_j)\vert} 
~ e^{-\frac{(\beta\sigma q_D)^2}{2}\left(\bnabla\phi({\bf x}_j) -  \bcalE({\bf x}_j)\right)^2} \Big. \nonumber \\
\hspace*{1cm} + \Big.2 \lambda_s  \cos{\beta e \phi({\bf x}_j)}~e^{-{\frac{(\beta\sigma q_I)^2}{2}\left(\bnabla\phi({\bf x}_j) -  \bcalE({\bf x}_j)\right)^2}} \Big). \nonumber \\ ~~~~~~~~~
\end{eqnarray}
One has $\lambda_D = e^{\mu_D}$ and $\lambda_S = e^{\mu_\pm}$ for an asymmetric electrolyte. Going to the continuum limit from here and assuming that all the sites have the same volume $a^3$ then the continuum limit of the above result is given by
\\
\begin{eqnarray}
\hspace*{-2.5cm} V[\bcalE({\bf x}), \phi({\bf x})] = 
\frac{1}{a^3} \int d^3{\bf r} 
\ln \Big(\lambda_D 
\frac{\sin\left(\beta q_D s~ \vert\bnabla\phi({\bf x}) - \bcalE({\bf x})\vert\right)}{\beta s~\vert\bnabla\phi({\bf x}) - \bcalE({\bf x})\vert}\Big) 
e^{-\frac{(\beta\sigma q_D)^2}{2} \left(\bnabla\phi({\bf x}) - \bcalE({\bf x})\right)^2}\Big.\nonumber \\
+ \Big. 2 \lambda_s  \cos{\beta e \phi({\bf x})} ~ 
e^{-\frac{(\beta\sigma q_I)^2}{2}
\left(\bnabla\phi({\bf x}) -  \bcalE({\bf x})\right)^2}\Big). 
\end{eqnarray}
\\
These are the one-body partition functions in the case of a mixture of the Drude oscillators and Drude electrolyte ions. Without the dipolar contribution the two expressions above reduce to the well-known Poisson-Boltzmann theory and the Poisson-Boltzmann lattice gas theory.

The form of these equations used in the main text, {\sl i.e.} Eqs. (\ref{bgwfxewa1}), is obtained by inserting the saddle point form (imaginary value) of the auxiliary field variables, which eventually converts the trigonometric  into hyperbolic functions and changes the sign of the Gaussian factors.  
\\ 

{\bf References}
\\

\end{document}